# Priming prosocial behavior and expectations in response to the Covid-19 pandemic - Evidence from an online experiment

Valeria Fanghella, Thi-Thanh-Tam Vu, Luigi Mittone[1]

**Abstract**

This paper studies whether and how differently projected information about the impact of the Covid-19 pandemic affects individuals' prosocial behavior and expectations on future outcomes. We conducted an online experiment with British participants (N=961) when the UK introduced its first lockdown and the outbreak was on its growing stage. Participants were primed with either the environmental or economic consequences (i.e., negative primes), or the environmental or economic benefits (i.e., positive primes) of the pandemic, or with neutral information. We measured priming effects on an incentivized take-and-give dictator game and on participants' expectations about future environmental quality and economic growth. Our results show that primes affect participants' expectations, but not their prosociality. In particular, participants primed with environmental consequences hold a more pessimistic view on future environmental quality, while those primed with economic benefits are more optimistic about future economic growth. Instead, the positive environmental prime and the negative economic prime do not influence expectations. Our results offer insights into how information affects behavior and expectations during the Covid-19 pandemic.

---

[1] Valeria Fanghella: University of Trento, Italy; Grenoble Ecole de Management, France. Thi-Thanh-Tam Vu: University of Trento, Italy; LUT University, Finland; Vietnam National University, Vietnam. Luigi Mittone: University of Trento, Italy; LUT University, Finland. Correspondence to: thithanhtam.vu@unitn.it

# 1    Introduction

The Covid-19 pandemic has far-reaching consequences beyond a health crisis. As an effort to contain the virus, governments have imposed restrictive measures including travel restrictions, social distance, stay-at-home order, and for the very first time, put the entire country under lockdown. On the one hand, the global restrictive measures may cause one of the worst economic recessions since the Great Recession.[2] The pandemic could cause nearly 200 million job losses,[3] and global economic growth may drop to half of the projected rate before the virus outbreak.[4] On the other hand, the environment has, at least in the short run, benefitted from the lockdown: air pollution and greenhouse gas emissions have temporarily dropped, and wildlife has taken over anthropized areas (Helm, 2020).

These unprecedented events have dramatically changed people's life and perception of the world, as well as the world itself. First, the dramatic changes caused by the disaster influence individuals' assessment of the future (Västfjäll, Peters and Slovic, 2008). Changes in one's expectation about the economy or the environment lead to ramifications in voting behavior (Lewis-Beck and Stegmaier, 2000), consumer confidence (Curtin, Presser and Singer, 2000), investment decision (Starr and Yilmaz, 2007), and willingness to mitigate climate change and other environmental issues (Chapman, Lickel and Markowitz, 2017; Hornsey and Fielding, 2016). Second, in the midst of tremendous amounts of uncertainty, confusion, and fear, individuals change how they think about and react to other people. They may act selfishly and even harm others as they perceive more threats and risks (van Bavel et al., 2020) or may behave altruistically as cooperation and norm-governed behaviors are common during the crisis time (Drury, 2018). Both of these opposite acts have been displayed in individuals' psychological and behavioral responses to the Covid-19 pandemic.[5]

Given the explosion of media coverage on the Covid-19 pandemic, individuals' prosocial behaviors and expectations can depend on the information to which they are exposed. Information can promote preventive behaviors and avoid pandemic-related panic (Lep, Babnik and Beyazoglu, 2020; van Bavel et al., 2020; West et al., 2020), thereby fostering altruistic rather than selfish behaviors and easing the management of the crisis. Moreover, how media covers environmental- and economic-related topics plays a vital role in the post-crisis responses. Whether environmental conservation will still be considered a priority depends on how the situation is addressed to the public (Henrique and Tschakert, 2019; Miles and Morse, 2007). If news raises excessive optimism about the environmental benefits of the pandemic, it will have a detrimental effect on conservation policies (Helm, 2020). On the other hand, news can speed up the economic recovery by raising public optimism and faith in it (Hollanders and Vliegenthart, 2011; van Giesen and Pieters, 2019). For these reasons, assessing the impact of news on individuals' prosociality and expectations is of high importance to manage a crisis like the Covid-19 pandemic and the post-crisis recovery.

---

[2] https://blogs.imf.org/2020/04/14/the-great-lockdown-worst-economic-downturn-since-the-great-depression/
[3] https://news.un.org/en/story/2020/04/1061322
[4] https://www.oecd.org/economic-outlook/
[5] See for example, https://www.theguardian.com/world/2020/mar/21/coronavirus-uk-panic-buyers-urged-to-think-of-frontline-workers; https://www.independent.co.uk/news/world/americas/coronavirus-supermarket-cough-prank-food-pennsylvania-hygiene-a9426781.html; https://www.theguardian.com/lifeandstyle/2020/mar/21/like-an-emotional-mexican-wave-how-coronavirus-kindness-makes-the-world-seem-smaller.

To unearth how individuals' prosociality and expectations are affected by the valenced information about the Covid-19 pandemic, we run an online experiment with British participants when the UK is under lockdown and the outbreak still in its growing phase. Participants are primed with different information about the impact of the pandemic that either emphasizes its benefits or consequences for the environment or the economy. We measure how the primes influence participants' prosociality through an incentivized take-and-give dictator game and their effect on expectations of environmental quality and economic growth.

Our results show that primes do not influence prosocial behavior. However, regardless of treatment assignment, participants behave very generously in the take-and-give dictator game, suggesting that the extraordinary time of the pandemic triggers altruism. Second, primes affect participants' expectations, and this effect varies across priming contents. Specifically, we find two main significant results: (1) negative environmental information induces pessimism about future environmental quality; (2) positive economic information triggers optimism about future economic growth. By contrast, the negative environmental information and the positive economic information do not influence expectations. We reason these different priming effects on the biased concentration of news coverage: the media has mostly provided one-dimensional information flow on the Covid-19 pandemic's environmental benefits and economic losses.[6] As such, only those primes that contain information not frequently covered by media (i.e., the pandemic's environmental consequences and economic benefits), significantly affect participants' expectations.

---

[6] See for example, https://ec.europa.eu/commission/presscorner/detail/en/ip_20_799 and https://www.imf.org/en/Publications/WEO/Issues/2020/04/14/weo-april-2020 for the economy; https://www.weforum.org/agenda/2020/04/coronavirus-covid19-air-pollution-enviroment-nature-lockdown and https://www.eea.europa.eu/highlights/air-pollution-goes-down-as for the environment.

The remainder of the paper proceeds as follows. Section 2 reviews relevant literature and draws the hypotheses of the study (pre-registered).[7] In Section 3, we present the experimental design and procedure followed by the experimental results in Section 4. Section 5 discusses the main results and offers some concluding remarks.

## 2 Hypothesis development

People's preferences depend on how information is presented (Tversky and Kahneman, 1985). In this study, we focus on a specific way to influence participants' behaviors and expectations: priming. Priming is a technique initially introduced by experimental psychologists and refers to the mental activation of a specific concept (Bargh and Chartrand, 2000). It has been widely adopted both in psychology and economics to influence participants' decisions (Cohn and Maréchal, 2016; Janiszewski and Wyer, 2014).

Our study belongs to the emergent literature of media priming. Media priming refers to how contents covered by the media influence people's behavior and decisions (Roskos-Ewoldsen, Roskos-Ewoldsen and Carpentier, 2009). Prior research has mainly studied priming intensity, recency, and valence of the media in various contexts including violence, politics, and stereotype (see Roskos-Ewoldsen et al., 2009 for a review). This body of studies does concur that by emphasizing certain aspects (e.g., definitions, solutions, consequences) of an issue and not others, media can influence people's attitudes and evaluations about that issue (Kalogeropoulos et al., 2017). According to Althaus and Kim (2006), media priming effects are the result of two mechanisms, namely accessibility and applicability of information. Accessibility is defined by the extent to which a specific knowledge can be activated while applicability refers to how "a stimulus and a stored knowledge construct are perceived as applicable to one another" (Althaus and Kim, 2006, p.962). Information emphasized by the media offers a shortcut in decision-making, because the primed content becomes particularly accessible and applicable when individuals have to make a decision related to the emphasis.

### 2.1 Priming and prosocial behavior

Priming has been erratically used in economics to promote prosocial behavior (Kamenica, 2012), especially in the context of natural disasters. More recently, a number of studies investigate how to foster compliance with measures to prevent the Covid-19 leveraging on social norms (Bilancini et al., 2020) or on moral frames (Everett et al., 2020). Yet, to the best of our knowledge, there has been no study that explicitly examined the priming effect of media on prosociality.

---

[7] Hypothesis and study pre-registration can be found at the following link: https://osf.io/td4r9/?view_only=18afc22e064647c0b45ceb84395b0ed6.

We employ the literature on how emotions affect prosociality to establish our first hypothesis. Positive or negative information can affect people's moods, and thus, prosocial behavior. Despite a large amount of research on the topic, the effect of negative mood on prosocial behavior is not conclusive. Some studies observed that people in a bad mood behave less prosocially than those in a good mood (e.g., Carlson and Miller, 1987; George, 1991), especially when the bad mood is caused by anger (Pillutla and Murnighan, 1996). Others found evidence that sad people compensate for their negative mood with good behavior (e.g., Cialdini, Baumann and Kenrick, 1981; Manucia, Baumann and Cialdini, 1984). Particularly, Hanley et al. (2017) failed to detect any effect of emotions on prosocial behavior. Given these inconsistent results, we explore the effect of positive and negative primes on prosociality, by studying which of the two alternative hypotheses holds in our setting:

*Hypothesis 1a: Participants exhibit higher (lower) prosociality after reading a paragraph that highlights the positive (negative) impact of the pandemic.*

*Hypothesis 1b: Participants exhibit lower (higher) prosociality after reading a paragraph that highlights the positive (negative) impact of the pandemic.*

Behavioral scientists and psychologists have shown that individuals are prone to loss aversion (also known as negativity bias) as losses result in more negative responses than gains of a similar magnitude result in positive responses (Tversky and Kahneman, 1974, 1991). In other words, people are more sensitive to losses (or negativity) than gains (or positivity). This phenomenon is also manifested with media priming. Previous studies on the negative and positive valence's priming effects (e.g., Bizer and Petty, 2005; Dillman Carpentier, Roskos-Ewoldsen and Roskos-Ewoldsen, 2008; Northup and Carpentier, 2013) found that people's judgment and behavior are more heavily influenced by negative than positive information (see Dijksterhuis, 2010 for a review). According to Sheafer (2007, p.23), negatively valenced stimulus induces a stronger effect than positively valenced one because it ''captures our attention far more than information about positive developments''. We also conjecture such an asymmetric effect in our setting, as stated in the following hypothesis:

*Hypothesis 2: The effect of priming on participants' prosociality is more pronounced when the prime is negative than when it is positive.*

In terms of priming environmental and economic concepts, we draw on a number of studies that investigate the effect of each of these concepts individually. The activation of environmental concepts is closely associated with the activation of prosocial concepts. Laboratory experiments show that priming with prosocial values increase contribution in a team contest (Andersson et al., 2017a) and charitable donation (Andersson et al., 2017b), but only among prosocial individuals. Drouvelis et al. (2015) show that activating the concept of cooperation increases voluntary public good provision. Similarly, Van der Werff et al. (2014) find evidence that recalling previous pro-environmental behaviors improves subsequent environmental judgments and intentions. This effect is maintained only when the information relates to positive environmental or sustainable concepts and not to negative ones. Notably, a degraded environment is more likely to trigger negative than positive behavior (Keizer, Lindenberg and Steg, 2008).

Economic concepts generally have a detrimental effect on prosocial behavior (Vohs, 2015). Adding a market institution increases self-interested and profit maximization behavior in public good games (Reeson and Tisdell, 2010). Similar findings have been observed in the context of environmental conservation: priming about ecosystem service valuation reduces donation to an environmental fund, compared to the condition when this information is not provided (Goff, Waring and Noblet, 2017). Activating economic concepts triggers a self-sufficient orientation and reduces requests for help and helpfulness toward others (Vohs, Mead and Goode, 2006).

Bridging together the two streams of literature, we expect that priming environmental concepts will lead to more prosocial behavior than economic ones, especially for positive priming. This leads to the following hypothesis:

*Hypothesis 3: Participants exhibit higher prosociality after reading a paragraph that highlights the positive environmental impact of the pandemic than the positive economic impact.*

To the best of our knowledge, there has been no previous study on the negative priming effect of environmental vs. economic concepts. Therefore, we explore which of the two following alternative hypotheses holds:

*Hypothesis 4a: Participants exhibit higher prosociality after reading a paragraph that highlights the negative environmental impact of the pandemic than the negative economic impact.*

*Hypothesis 4b: Participants exhibit lower prosociality after reading a paragraph that highlights the negative environmental impact of the pandemic than the negative economic impact.*

## 2.2 Priming and expectations

Natural and anthropogenic disasters like the Covid-19 pandemic affect people's perceptions and attitudes on future outcomes (Västfjäll et al., 2008). With regard to environmental issues, being exposed to natural disasters affects people's evaluation of risks and of mitigation and adaptation measures. The Fukushima nuclear accident increased the Chinese public's risk perception of nuclear disasters and significantly reduced nuclear plants acceptability (Lei et al., 2013). Media plays a role inasmuch as it shapes which consequences of the disaster are brought to public attention (Miles and Morse, 2007). Moreover, whether media raises more or less support for policies for environmental risk prevention depends on how it frames the information (Sunstein, 2007). Indeed, availability bias and information framing affect attitude towards natural disasters (Rheinberger and Treich, 2017).

There is a growing body of work suggesting that positive and negative information influences individuals' economic evaluation (e.g., Garz, 2013; Petalas, Van Schie and Vettehen, 2017; Soroka, 2006). Using a laboratory experiment, Bosman, Kräussl and Mirgorodskaya (2017) examined if investors' expectations are influenced by the valence of financial news they read. Their findings suggest that people predict higher stock price turn, perceive less risk, are more likely to buy and especially, hold a more optimistic view on the economy after reading positive than negative news. Consistently, Hollanders and Vliegenthart (2011) found that negative news coverage on economic development during the period 1990-2009 was correlated with lower consumer confidence and this effect was more pronounced at the beginning of the credit crisis. Kalogeropoulos et al. (2017) expanded these findings, by showing that positive and negative news about the economy not only affects expectations about economic development but also on overall Government evaluation. In line with the priming effect on prosocial behavior, the asymmetric responsiveness of people to positive and negative information is also explored in these studies: participants are more sensitive to negative than positive information (Bosman, Kräussl and Mirgorodskaya, 2017; Petalas et al., 2017).

Based on the above findings, we hypothesize that positive and negative priming respectively increases optimism and pessimism in individuals' expectations about economic and environmental outcomes, and that the negative one has a greater impact than the positive one.

*Hypothesis 5: Participants expect a higher (lower) future economic growth and environmental quality after reading a paragraph that highlights the positive (negative) impact of the pandemic.*

*Hypothesis 6: The effect of priming on participants' expectations on future outcomes is more pronounced when the prime is negative than when it is positive.*

The accessibility mechanism of priming suggests that participants rely on information about a specific issue to make their judgment related to that issue. Janiszewski and Wyer (2014) provided a thorough review on content priming and summarized that "an increase in the accessibility of content increases the likelihood the content will be integrated into ongoing perceptions, judgements, and choices" (p.97). As an example, Kalogeropoulos et al. (2017) found that individuals rely on economic news to form their economic evaluations. This leads us to the last hypothesis:

*Hypothesis 7: Economic (environmental) priming influences future economic (environmental) outcome predictions.*[8]

## 3    Experimental design

### 3.1    Stages

#### 3.1.1  Stage 1: Take-and-give dictator game

In the first stage, participants played a simplified version of the take-and-give dictator game (Cappelen et al., 2013; List, 2007). In this version of the dictator game, also negative transfers are allowed. Each participant was randomly matched with another participant and was assigned the role of dictator or recipient. In a pair, the dictator and the recipient were given 5 pounds each. The dictator selected, from a list in steps of 1 pound, how much to transfer to the recipient between -5 and +5 pounds. A positive amount meant that the dictator gives the recipient part of her money. A negative amount that the dictator took from the budget of the recipient. The order of presentation of choices (descending vs. ascending) was randomized.

---

[8] In the pre-registration, we initially had two hypotheses for Hypothesis 7. *H1: Participants tend to expect a higher effect on economic than environmental outcomes after reading a paragraph that highlights the economic impact of the pandemic. H2: Participants tend to expect a higher effect on environmental than economic outcomes after reading a paragraph that highlights the environmental impact of the pandemic.* We collapsed them in one for clarity's and conciseness' sake, but the meaning remains unchanged.

All participants made the decision in the role of dictators. At the end of the experiment, we randomly selected 5% of participants as dictators, 5% as recipients, and we matched them. They were paid according to the dictator's allocation.

The take-and-give dictator game was employed to measure participants' pro and antisocial behavior in response to the primes. We introduced the take-and-give version of the dictator game because it captured competitive behaviors towards others in a context of resource scarcity (e.g., stockpiling protective masks and durable food without leaving enough for other people).

### 3.1.2 Stage 2: Outcome predictions

In the second stage, participants were asked to make predictions on how the pandemic may affect economic growth and environmental quality for the current year. We used two indicators of economic growth and environmental quality, which were the gross domestic product (GDP) growth rate and the density of particulate matter (PM), respectively. We presented four scenarios about GDP growth rate and PM density for Europe and worldwide. Two scenarios were optimistic and the other two were pessimistic. The degree of optimism/pessimism varied across scenarios. We constructed the four scenarios in a way that increased their realism. Notably, we based them on previous years' data and experts' predictions on possible post-pandemic developments.

For the pessimistic economic scenarios, we used the predictions on how the pandemic affected the European and global economy presented by experts of McKinsey & Company in their Covid-19 briefing notes of the 4th of April.[9] These were two negative scenarios that varied the degree of severity on Europe and worldwide (henceforth, *pessimistic* and *highly pessimistic*). We constructed the *optimistic* scenario using predictions on the real GDP growth rate in 2020 made before the outbreak. For the *highly optimistic* scenario, we employed the highest GDP growth rate for Europe and worldwide since the 2008 crisis.[10] Table 1 reports the economic scenarios.

---

[9] https://www.mckinsey.com/business-functions/risk/our-insights/covid-19-implications-for-business
[10] https://www.imf.org/external/datamapper/NGDP_RPCH@WEO/OEMDC/ADVEC/WEOWORLD

**Table 1. Economic scenarios, from the most pessimistic to the most optimistic**

|            | Highly pessimistic | Pessimistic | Optimistic | Highly optimistic |
|------------|--------------------|-------------|------------|-------------------|
| Europe (%) | -10.6              | -4.7        | +1.7       | +4.6              |
| World (%)  | -5.7               | -1.8        | +3.4       | +5.6              |

For the environmental outcomes, we could not find the same predictions as for the economy about the effect of the pandemic. As a reference point, we used the temporary effect that the lockdown had on PM2.5 density in China, equal to a reduction of 20-30%.[11] The *optimistic* scenario envisaged a reduction of 10% in PM density and the *highly optimistic* one of 30%. We also used these percentages to create the *pessimistic* (+10%) and *highly pessimistic* scenarios (+30%). We provided participants with scenarios about absolute values of PM density, not as percentages of change. Notably, we applied the percentages to the latest available data of PM2.5 density before the crisis: 19 µg/m3 and 46 µg/m3, for Europe and worldwide, respectively.[12] Table 2 reports the environmental scenarios.

**Table 2. Environmental scenarios, from the most pessimistic to the most optimistic**

|                 | Highly pessimistic | Pessimistic | Optimistic | Highly optimistic |
|-----------------|--------------------|-------------|------------|-------------------|
| Europe (µg/m3)  | 24.7               | 20.9        | 17.1       | 13.3              |
| World (µg/m3)   | 59.8               | 50.6        | 41.4       | 32.3              |

Participants were presented with Table 1 and 2. Their order, as well as that of scenarios, were randomized. Besides, we also provided the participants with the latest available values for the European and Global indicators (summarized in Table 3).[13]

**Table 3. Latest available information on economic and environmental indicators**

|        | GDP growth rate (2019) | PM density (2017) |
|--------|------------------------|-------------------|
| Europe | +1.5%                  | 19 µg/m3          |
| World  | +3%                    | 46 µg/m3          |

Before participants made their predictions, they were provided with some background information. We explained the meaning and the interpretation of the indicators included in this stage. We also highlighted whether a positive or negative change in these indexes could be considered as an improvement for the economy and the environment. Participants could proceed to the predictions only if they answered correctly the two following instructional checks: "Is the growth of the gross domestic product (GDP) good or bad for the economy of a country?" and "Is the increase in density of particulate matter (PM) good or bad for the environment?".

---

[11] https://atmosphere.copernicus.eu/amid-coronavirus-outbreak-copernicus-monitors-reduction-particulate-matter-pm25-over-china
[12] https://www.stateofglobalair.org/data/#/air/plot
[13] PM density was taken from https://www.stateofglobalair.org/data/#/air/plot while the real GDP growth rate was taken from https://www.imf.org/external/datamapper/NGDP_RPCH@WEO/OEMDC/ADVEC/WEOWORLD.

This stage aimed to examine whether our manipulations affect participants' degree of optimism and pessimism. Even if these predictions were central to our research question, we could not incentivize this stage. It should be noted that we conducted our study in the time of change so we could only adopt the most likely scenarios of future outcomes as projected by experts by the time of the experiment.

### 3.1.3 Stage 3: Questionnaires

The third stage contained a questionnaire about participants' social and risk preferences and other Covid-19 related questions, such as the degree of concern and of acceptance of governmental policies.

We elicited social preferences with the version of the Social Value Orientation (SVO)[14] developed by Murphy et al. (2011). Participants were asked to make six choices on how to allocate money between themselves and a hypothetical partner. Based on their preferred allocations, participants were classified into one of four types: (1) altruistic (i.e., individuals who aim at maximizing others' payoffs); (2) prosocial (i.e., individuals who aim at maximizing the joint payoffs); (3) individualistic (i.e., individuals who aim at maximizing their own payoffs); (4) competitive (i.e., individuals who aim at maximizing the positive difference between their own and others' payoffs).

We elicited risk preferences using the Bomb Risk Elicitation Task (BRET) (Crosetto and Filippin, 2013). Participants were asked to choose the number of boxes out of 64 that they want to open given that one box contains the bomb. The hypothetical payment increased linearly with the number of boxes collected but becomes zero if the bomb is also collected. Both SVO task and BRET were not monetarily incentivized.

This stage also includes an attention check to screen participants who did not pay attention while completing the experiment. For this check, participants were asked to choose exactly "Strongly disagree" to a question.

---

[14] *SVO* is defined as a construct that theoretically extends the rational self-interest assumption in traditional economics theories by assuming that economic agents have broader motivation such as maximizing joint outcomes, enhancing equality in outcomes or minimizing others' outcomes (Van Lange, 1999).

### 3.2 Treatments

At the beginning of the experiment, participants were randomly assigned to one of the five different experimental treatments. Two treatments highlighted the positive effects of the pandemic on the environment, *Env+*, and on the economy, *Econ+*. The other two treatments, *Env-* and *Econ-*, stressed the environmental and economic consequences of the outbreak, respectively. We also introduced a baseline condition named *Neutral*, where we provided information about the effect of reading books. The baseline condition allows us to test our hypotheses that negative priming induces greater effect than positive priming. Namely, simply contrasting positive vs. negative primes does not distinguish the contribution of each priming to any difference in the outcome variables. The baseline condition provides a benchmark against which to contrast the effect of each priming and determine the one with largest effect.

In each treatment, there was a short paragraph displayed at the beginning of the experiment and at the beginning of each stage. The first priming was given after the participants consented to join the experiment, followed by three control questions to ensure that participants read this priming. The other two primes (i.e., two other paragraphs) were displayed before participants made decisions in Stage 1 and Stage 2. Each paragraph reported some plausible impact of the pandemic on the environment or the economy. Table 4 summarizes the description of each treatment and an example of a priming paragraph.

### 3.3 Experimental procedures

We conducted the experiment on the online platform Prolific using oTree (Chen, Schonger and Wickens, 2016). Online platforms offer reliable results for experimental research, as shown by replications of well-known experiments (Horton, Rand and Zeckhauser, 2011; Palan and Schitter, 2018; Paolacci, Chandler and Ipeirotis, 2010). Moreover, they also allow for better external validity when compared to research conducted using student samples (Henrich, Heine and Norenzayan, 2010). We ran four experimental sessions between the 15 and the 23 of April 2020. To be eligible, participants had to be UK citizens and live in the UK at the time of the experiment. It should be noted that at the time we ran the experiment, the entire UK was under its first lockdown and the Covid-19 outbreak was on its growing stage.[15]

---

[15] https://www.worldometers.info/coronavirus/country/uk/

**Table 4. Summary of the treatments**

| Treatment | Priming features | | Example of priming paragraph |
|---|---|---|---|
| | Valence | Concept | |
| Env+ | Positive | Environmental | Covid-19 pandemic has some positive effects on the environment. Amongst the others, it may increase the usage of sustainable working practices such as online conferences and digital libraries. |
| Econ+ | Positive | Economic | Covid-19 pandemic has some positive effects on the economy. Amongst the others, it may increase the development of cost-effective innovations such as digitally-enabled forms and smart-working tools. |
| Env- | Negative | Environmental | Covid-19 pandemic has some negative effects on the environment. Amongst the others, it may decrease the allocation of resources on pro-environmental policies such as subsidies for green energy. |
| Econ- | Negative | Economic | Covid-19 pandemic has some negative effects on the economy. Amongst the others, it may decrease the number of small and medium businesses in some sectors such as tourism and transportation. |
| Neutral | Neutral | Neutral | Reading books has some effects on the mind. Amongst the others, it may affect writing capabilities, such as the usage of punctuation marks and of rare words. |

The experiment started with a paragraph about the Covid-19 outbreak for all participants, regardless of treatment assignment, to make them aware that the experiment was related to the pandemic and to provide them with the same basic knowledge about it and its health consequences. Participants proceeded to the first priming paragraph, whose content depended on their treatment. They were then instructed on the take-and-give dictator game of Stage 1. Before choosing how much to transfer, participants were exposed to the second priming paragraph. After the instructions of Stage 2, we primed participants with the last paragraph and we asked their predictions about environmental and economic outcomes. Finally, participants completed the questionnaire in Stage 3.

Overall, 988 participants completed the study. We excluded 22 participants who failed the attention check and 2 participants who completed the study too quickly (less than 4 minutes), and 3 participants who had missing socio-demographic variables. The final sample includes 961 participants. On average, participants completed the experiment in 11 minutes. Each participant was given 1.8 pounds for participation. In addition, we randomly selected 10% of participants for payment in the take-and-give dictator game, 5% as dictators and 5% as recipients.

While running the first two experimental sessions, we made a typo in the table with environmental outcomes. We corrected it on the two final sessions. Therefore, we analyzed the hypotheses on environmental predictions using a subsample of 437 participants. We excluded that the problem also affected the other outcomes of the study. First, the take-and-give dictator game took place before the outcome predictions and is thus unaffected. Second, economic predictions were presented on a different table.

## 4    Results

We first summarize demographic characteristics of participants across treatments and then focus on the most substantial research issues of our study, which are the priming effect on individuals' prosociality, environmental quality and economic growth expectations.

**Table 5. Participants' characteristics**

|  | Env+ | Econ+ | Env- | Econ- | Neutral |
|---|---|---|---|---|---|
| N | 180 | 190 | 202 | 192 | 197 |
| Male (%) | 26.11 | 27.37 | 23.27 | 28.65 | 24.87 |
| Age (mean) | 33.48 | 34.11 | 34.03 | 34.10 | 33.32 |
|  | (11.82) | (12.57) | (11.24) | (12.05) | (10.94) |
| University degree (%) | 0.49 | 0.51 | 0.53 | 0.46 | 0.53 |
| BRET (mean) | 22.67 | 22.39 | 22.00 | 22.14 | 21.54 |
|  | (15.15) | (15.04) | (13.67) | (13.52) | (13.90) |
| SVO |  |  |  |  |  |
| Competitive (%) | 1.67 | 1.50 | 0.50 | 0.52 | 2.54 |
| Individualistic (%) | 17.22 | 13.68 | 14.36 | 17.19 | 21.83 |
| Prosocial (%) | 79.44 | 85.26 | 84.16 | 80.21 | 74.62 |
| Altruistic (%) | 1.67 | 0.00 | 0.99 | 2.08 | 1.02 |

Note: Standard deviations in parentheses

Table 5 reports participants' characteristics across treatments. All socio-demographic variables are balanced across experimental conditions. Overall, 26% of participants are male, and 50.6% have at least an undergraduate degree. Age ranges between 18 and 78 years. Risk attitudes measured by the BRET and social value orientation measured by the SVO task are uniformly distributed across treatments. On average, participants open 20 out of 64 boxes. The percentage of prosocial and individualistic types are 81% and 17% respectively. Each of the other two types (competitive and altruistic) accounts for less than 2% of the sample.

### 4.1 Priming and prosociality

On average, participants transferred 1.18 pounds to another anonymous participant. Figure 1 illustrates the distribution of participants' decisions in the take-and-give dictator game across treatments. In all treatments, more than 20% of participants did not transfer anything; the median of the transfer is 2 pounds. About 5% chose to take all 5 pounds from the endowment of the other partner while more than 60% transferred a positive amount. These results show more altruistic behaviors than previous studies using the same game (e.g., Cappelen et al., 2013; List, 2007). As an example, List (2007), in his treatment named "Take ($5)" which is the most similar to our experimental setting, found that only 10% of participants transfer a positive amount.

**Figure 1.** Distribution of transfers in the take-and-give dictator game across treatments

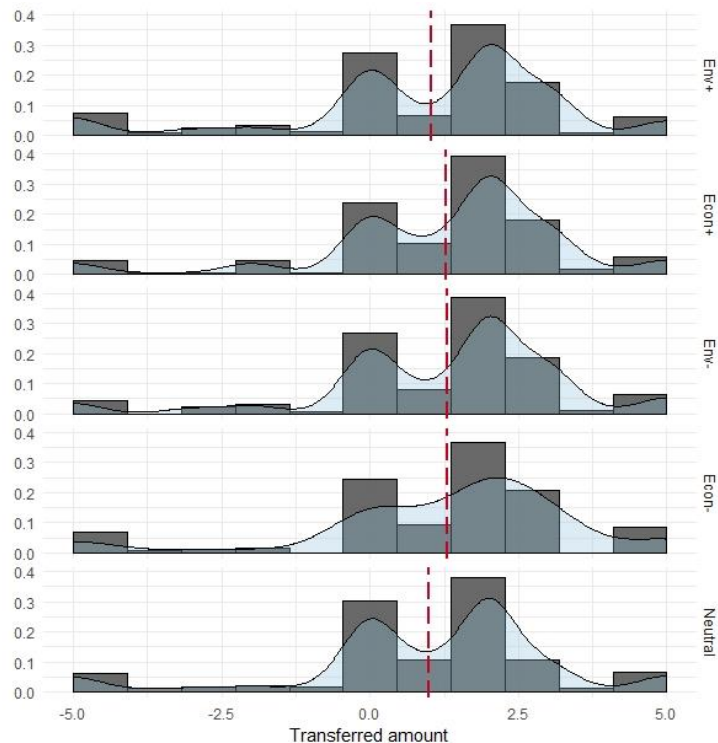

Note: Red dashed line represents mean transfer per experimental condition

We first examine Hypothesis 1 and 2 on the effect of positive vs. negative prime. For positive primes, we pool the observations of the *Env+* and *Econ+* treatment, while for the negative primes, those of the

*Env-* and *Econ-*. Table 6 reports the results from the OLS regressions, where the dependent variable is participants' transfer, and the regressors are treatments (M1). In model M2, we also include measures of risk preferences (denoted as BRET), SVO, and other demographic characteristics such as gender, age, and whether they have a university degree as independent variables.

**Table 6. Effect of positive and negative prime on transfers in the take-and-give dictator game**

|  | M1 |  | M2 |  |
| --- | --- | --- | --- | --- |
|  | β | SE | β | SE |
| Negative | 0.14 | (0.16) | 0.10 | (0.15) |
| Neutral | -0.16 | (0.19) | -0.05 | (0.19) |
| BRET |  |  | -0.00 | (0.00) |
| Competitive |  |  | -2.66*** | (0.61) |
| Individualistic |  |  | -1.25*** | (0.18) |
| Altruistic |  |  | 1.51** | (0.64) |
| Female |  |  | 0.52*** | (0.15) |
| Age |  |  | 0.02*** | (0.01) |
| University degree |  |  | -0.17 | (0.14) |
| Constant | 1.16*** | (0.11) | 0.65** | (0.29) |
| Observations | 961 |  | 961 |  |
| R-squared | 0.00 |  | 0.09 |  |

Note: OLS regression, standard errors in parentheses. ***$p < 0.01$, **$p < 0.05$, *$p < 0.1$. *Negative* denotes the negative prime (*Env-* and *Econ-* together), *Neutral* denotes the neutral prime. *Positive (Env+* and *Econ+* together) is the reference group. BRET represents risk aversion. Competitive, Individualistic and Altruistic are SVO types. Prosocial is the SVO reference group.

Regression results suggest that there is no priming effect on participants' prosocial behavior. The coefficient of the negative primes does not significantly differ from 0 in both specifications, showing that the average transfers between the two primings are not statistically different. The lack of significant effect is not unexpected: we formulated Hypothesis 1 as exploratory, as previous findings on the topic are inconclusive.

Next, we investigate Hypothesis 2 on the asymmetric effect between positive and negative primes. Using F-tests, we test whether transfers with positive and negative primes significantly differ from the one in the *Neutral* condition. If Hypothesis 2 holds, the difference between the negative and the neutral prime should be higher than that between the positive and the neutral primes. However, F-tests show that there is no difference between any of these primes. Hence, we find no empirical support for Hypothesis 2.

We then test Hypothesis 3 and 4 by assessing whether there is any between-treatment difference. As before, we run OLS regressions on participants' transfers without (M3) and with (M4) demographic characteristics. Table 7 shows no significant difference between any priming treatment and the baseline condition. As a direct test of our hypotheses, we first investigate whether participants behave more prosocially after being exposed to positive environmental priming than positive economic priming (Hypothesis 3). F-tests show that participants in the *Env+* and *Econ+* treatment transfer the same amount of money. Hypothesis 3 is thus not supported. Next, we perform the same analysis on the *Env-* and *Econ-* treatment. No significant difference is observed, providing no support to any of the two alternative formulations of Hypothesis 4.[16]

**Table 7. Effect of treatments on transfers in the take-and-give dictator game**

|  | M3 β | SE | M4 β | SE |
|---|---|---|---|---|
| Env+ | 0.02 | (0.23) | -0.07 | (0.22) |
| Econ+ | 0.29 | (0.22) | 0.17 | (0.21) |
| Env- | 0.30 | (0.22) | 0.13 | (0.21) |
| Econ- | 0.31 | (0.22) | 0.18 | (0.21) |
| BRET |  |  | -0.00 | (0.00) |
| Competitive |  |  | -2.64*** | (0.61) |
| Individualistic |  |  | -1.24*** | (0.18) |
| Altruistic |  |  | 1.54** | (0.64) |
| Female |  |  | 0.53*** | (0.15) |
| Age |  |  | 0.02*** | (0.01) |
| University graduate |  |  | -0.17 | (0.14) |
| Constant | 1.00*** | (0.16) | 0.60** | (0.30) |
| Observations | 961 |  | 961 |  |
| R-squared | 0.00 |  | 0.09 |  |

Note: OLS regression, standard errors in parentheses. ***p < 0.01, **p < 0.05, *p < 0.1. *Env+* and *Econ+* respectively denote the priming with positive environmental information and with positive economic information. *Env-* and *Econ-* respectively denote the priming with negative environmental and with negative economic information. *Neutral* is the reference group. BRET represents risk aversion. Competitive, Individualistic and Altruistic are SVO types. Prosocial is the SVO reference group.

To sum up, we do not find any significant priming effect on prosociality. However, as highlighted at the beginning of this section, participants are very generous in our experiment. The very high propensity of the dictators to donate to their recipients is one of the most important results that emerge from this study. Unfortunately, the considerations that can be drawn from this finding are mainly speculative because we did not formulate an ex-ante hypothesis about it. We explain dictators' high propensity to "donate" to their recipients with the psychological impact generated by Covid-19. In fact, it is commonly acknowledged in the dictator game literature that many factors explain the decision to transfer money to the recipients, and, among them, psychological mechanisms play an important role (Bardsley 2008).

---

[16] As a further check, we examine the determinants of transferring a positive amount by running logit regressions, where the dependent variable is equal to 1 if the transfer is strictly positive, and 0 otherwise. Results of the logistic regression are available upon request and replicate the lack of treatment effect on prosociality.

Participants' generosity during the Covid-19 outbreak is consistent with the social pressure model (Akerlof and Kranton, 2000). This model posits that the decision to behave altruistically is triggered by a demand external to the donor's will, in some way expressed by the society. Accordingly, this theoretical view states that what matters in triggering altruistic behavior is the perception of what the society expects from one as a society member. The social values prevailing in a society can be strongly modified when extraordinary historical events occur. As a consequence of these changes also individual behaviors can change (Cassar, Healy and von Kessler, 2017; Whitt and Wilson, 2007). This is certainly the case of the crisis generated by the Covid-19 pandemic, which triggered a demand for collaborative behavior and support for the needs of the most fragile members of the society. Our experimental results would seem to support the existence of this mechanism of indirect conditioning of individual behaviors, triggered by a social demand for help that, in some way, influences the decisions of the participants in the dictator game, pushing them towards more altruistic choices.

## 4.2 Priming and environmental quality predictions

The distributions of participants' predictions about 2020 particulate matter (PM) density of Europe and of the world are respectively depicted in Figure 2. For both regions, participants hold optimistic expectations on future environmental quality: except in the *Env-* treatment, the most frequently chosen scenario is the *highly optimistic* one, which entails a reduction of 30% in the PM density. Less than 25% of participants selected one of the two pessimistic scenarios for both European and global PM density.

**Figure 2. Participants' predictions on European (Panel A) and global (Panel B) environmental quality**

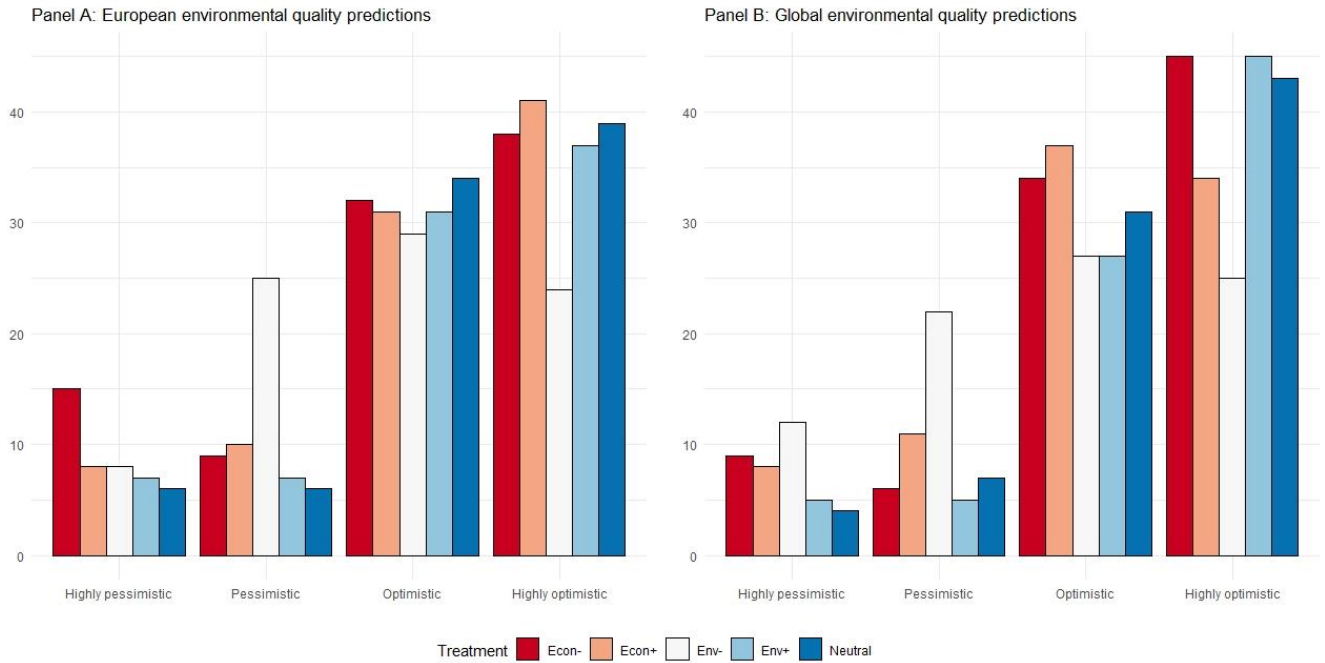

We first compare the impact of priming on predictions between positive and negative primes and then, across treatments. Before we present the results, it should be recalled that Hypothesis 5 states that positive and negative primings prompt optimism and pessimism, respectively. The effect of the negative priming is expected to be higher than that of the positive (Hypothesis 6). Finally, we expect that the effect of the priming is stronger on the indicator that shares the same concept with the priming (Hypothesis 7).

We assess treatment effect using the two-sided Fisher's exact tests. As we compare more than two treatments for each outcome (List, Shaikh and Xu, 2019), we correct the p-values for multiple hypothesis testing using Holm's correction.[17]

Table 8 tabulates the p-values from the Fisher's exact tests on the difference between positive and negative primes, and neutral information. For the European scenarios, participants' expectations significantly differ between the positive and negative primes. In line with Hypothesis 5, Table 8 shows that participants in the positive prime are more optimistic than those in the negative one. This difference is mostly driven by the negative priming: Participants in the positive priming have the same expectations than those in the neutral condition, whereas those in the negative priming significantly differ from it. Hence, Hypothesis 6 is also supported. The pattern of results is qualitatively, but not quantitatively, replicated for global outcomes.

---

[17] Results without Holm's correction are available upon request. Obviously, report more significant differences between pairs than the corrected tests. The comparisons that are not robust to the multiple hypothesis correction are those significant at 0.10, while those at 0.05 are generally robust.

**Table 8. Comparison of positive and negative priming effect on environmental quality predictions**

| Comparison | European | Global |
|---|---|---|
| Positive vs. Negative priming | 0.06 | 0.27 |
| Positive vs. Neutral priming | 0.85 | 0.82 |
| Negative vs. Neutral priming | 0.06 | 0.19 |

Note: p-values from two-sided Fisher's exact test. For each outcome, p-values are corrected for multiple hypothesis testing using Holm's correction. *Negative* denotes the negative prime (*Env-* and *Econ-* together), Positive the positive prime (*Env+* and *Econ+* together), *Neutral* denotes the neutral prime.

We further unpack these findings by assessing between-treatment differences. Table 9 reports the comparisons that are relevant to our hypothesis testing. Compared to participants exposed to the neutral information (*Neutral*) or to the economic consequences of the pandemic (*Econ-*), those primed with environmental consequences (*Env-*) are significantly more likely to choose the pessimistic scenarios. Surprisingly, *Env+* does not yield the same results, leading to the same expectations as the economic and neutral primes. Finally, as expected, priming with economic concept does not affect expectations about environmental outcomes. Overall, these results show that environmental priming influences environmental predictions more than what economic priming does (but only for negative information). Hence, for the environment, we partially confirm the effect of priming concept as conjectured in Hypothesis 7.

**Table 9. Comparison of treatment effect on environmental quality predictions**

| Comparison | European | Global |
|---|---|---|
| Env+ vs. Econ+ | 1.00 | 0.54 |
| Env+ vs. Neutral | 1.00 | 1.00 |
| Env- vs. Econ- | 0.03 | 0.01 |
| Env- vs. Neutral | 0.01 | 0.01 |
| Econ+ vs. Neutral | 1.00 | 0.94 |
| Econ- vs. Neutral | 1.00 | 1.00 |

## 4.3 Priming and economic growth predictions

The distributions of participants' predictions about 2020 GDP growth of Europe and of the world are depicted in Figure 3. In all treatments but the *Econ+*, more than 70% of participants expect a negative GDP growth rate in Europe and worldwide. The distribution between the two pessimistic scenarios differs slightly between the two regions. In particular, the most frequently chosen scenario is the *pessimistic* one for Europe (growth rate of -4.7%) and the *highly pessimistic* one for the world (growth rate of -5.7%). The proportion of participants selecting a negative GDP growth rate is the lowest in the positive economic priming (*Econ+*) with 57%. Overall, the pessimistic expectations about economic growth sharply contrast the optimism observed in the environmental predictions.

**Figure 3. Participants' predictions on European (Panel A) and global (Panel B) economic growth**

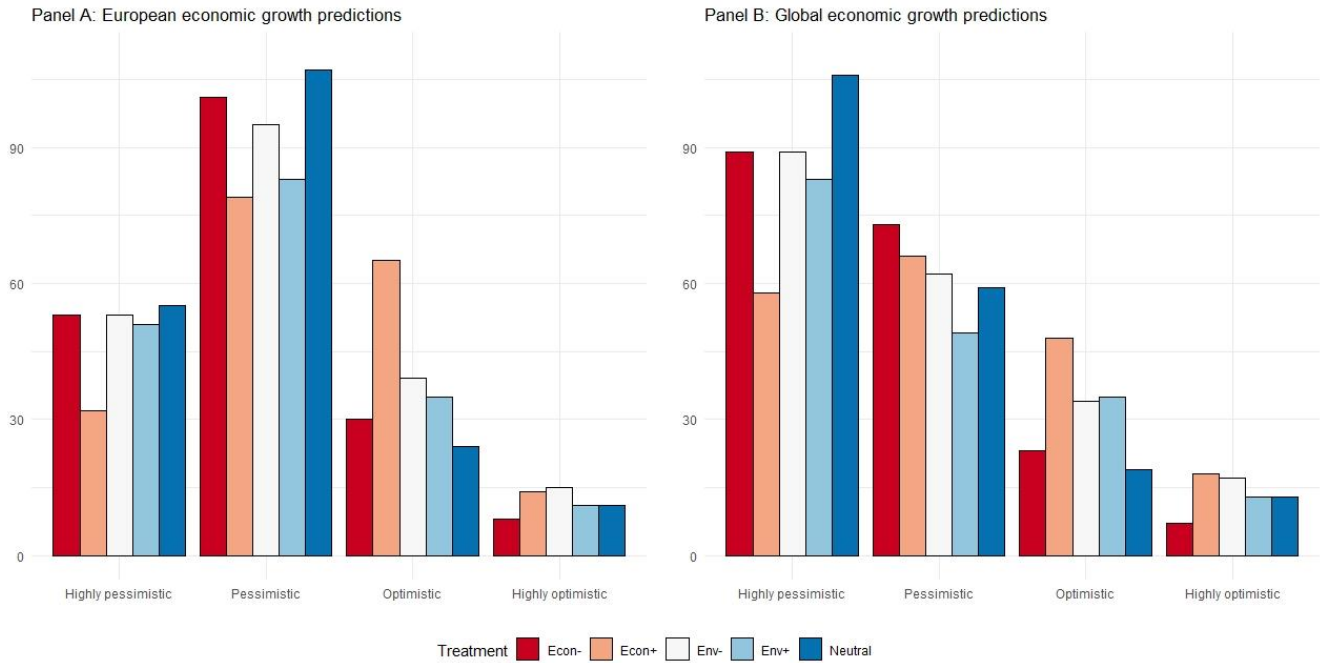

We repeat our hypothesis testing on economic predictions as for the environmental outcomes. In Table 10, we report the effect of priming positive vs. negative vs. neutral information. Results show that participants primed with positive information about the pandemic are significantly more optimistic about both European and global economic growth than their counterparts primed with negative or neutral information. These findings provide further support for Hypothesis 5.

Next, we find no significant difference in the distribution of GDP growth prediction between negative and neutral priming. Instead, positive priming triggers significantly more optimism than the neutral one. This is in sharp contrast with Hypothesis 6 and with the results of the environmental primes, in which we predicted and found that that negative prime has the largest effect.

**Table 10. Comparison of positive and negative priming effect on economic growth predictions**

| Comparison | European | Global |
|---|---|---|
| Positive vs. Negative priming | 0.02 | 0.02 |
| Positive vs. Neutral priming | 0.00 | 0.00 |
| Negative vs. Neutral priming | 0.40 | 0.16 |

Note: p-values from two-sided Fisher's exact test. For each outcome, p-values are corrected for multiple hypothesis testing using Holm's correction. *Negative* denotes the negative prime (*Env-* and *Econ-* together), Positive the positive prime (*Env+* and *Econ+* together), *Neutral* denotes the neutral prime.

Table 11 reports between-treatment differences. Results are in line with the differences across positive, negative, and neutral priming. Positive economic information (*Econ+*) leads to more optimistic predictions on the European economy than the positive environmental (*Env+*) and neutral priming. In addition, in the *Econ+* treatment, participants are significantly more optimistic about the global economy than in the *Neutral* treatment. Instead, participants in *Econ-* make the same predictions as those in the other treatments. This asymmetric result is again in contrast with our predictions since we hypothesized that economic primes, regardless of their valence, have higher effect on economic expectations than environmental primes (Hypothesis 7). However, it is consistent with the findings from the environmental outcomes, but for the opposite valence. Notably, the effect of the economic information is significant only when it is positive, whereas the effect of the environmental information is significant only when it is negative. In Section 5, we discuss how these seemingly contrasting outcomes may be derived from the same mechanism.

**Table 11. Comparison of treatment effect on economic growth predictions**

| Comparison | European | Global |
|---|---|---|
| Env+ vs. Econ+ | 0.02 | 0.12 |
| Env+ vs. Neutral | 0.72 | 0.20 |
| Env- vs. Econ- | 0.73 | 0.22 |
| Env- vs. Neutral | 0.72 | 0.22 |
| Econ+ vs. Neutral | 0.00 | 0.00 |
| Econ- vs. Neutral | 0.74 | 0.22 |

Note: p-values from two-sided Fisher's exact test. For each outcome, p-values are corrected for multiple hypothesis testing using Holm's correction. *Env+* and *Econ+* respectively denote the priming with positive environmental information and with positive economic information. *Env-* and *Econ-* respectively denote the priming with negative environmental and with negative economic information. *Neutral* denotes the neutral prime.

## 5 Discussion and conclusion

The goal of this study is to investigate how information, through priming, affects people's prosocial behavior and expectations amidst a global disaster, more specifically, the Covid-19 pandemic. To do so, we conduct an online experiment with British participants during the first UK lockdown. To reproduce the effect that news has on one's behavior and judgment, we expose participants to different kinds of information, which is either neutral or addresses the positive or the negative impact of the pandemic on the environment or the economy. We assess the effect of this information on participants' prosocial behavior through an incentivized take-and-give dictator game. We also investigate the effect on participants' expectations about future outcomes by asking their predictions on how the environment and the economy will be affected by the pandemic.

Our results are partially consistent with the existing literature. We observe that none of the primes influences prosocial behavior. Prior research found that negative or positive stimuli, as well as prosocial appeals, do not affect prosociality (Andersson et al., 2017a; Andersson et al., 2017b; Hanley et al., 2017). We show that this result holds also in the extraordinary context of the Covid-19 pandemic. Nevertheless, we fail to replicate a common result as discussed in Vohs (2015): in our setting, economic information does not prompt selfish behavior. An interesting outcome of our study is the high degree of altruism displayed by the participants regardless of the type of priming. This result is likely to be driven by the specific context of our experiment. Prior evidence has shown that sharing a traumatic event increases prosocial behavior, such as in-group cooperation (Whitt and Wilson, 2007) and trust (Cassar et al., 2017). Furthermore, the onset of demand from society in favor of the vulnerable citizens seemed to fuel the individual attitude towards altruistic behavior in general. This finding is of practical relevance for policy makers who target individuals' behavior to contrast the diffusion of the pandemic.

Second, we find that highlighting specific consequences of the pandemic influences how people expect the situation to develop. In terms of environmental expectations, the negative prime triggers more pessimism than the positive one. This is stronger when the negative prime contains information about the environment. Instead, the positive environmental priming has no effect on participants' environmental expectations. For the economic expectations, we observe the opposite pattern: the positive prime increases optimism in the economy compared to the other primes. On the contrary, the negative priming, even when it conveys negative economic information, does not affect participants' economic expectations. The fact that the positive prime has greater effect than the negative one contradicts decades of literature showing that people are more responsive to negative than positive information.

A search on Google News shows that since the virus first struck in China, media have productively covered the two inevitable impacts of the global shutdown: economic damage and environmental benefit. Such external priming may have generated a sort of ceiling effect on participants' expectations. More specifically, questioned about environmental quality predictions, people are, by default, optimistic. Only negative priming – information about environmental consequences of the pandemic – differs from what people already know and thus, changes their expectations about future environmental quality. Similarly, given the intensity of news on on-going economic consequences, participants already hold a pessimistic view on how the pandemic will affect economic development. Hence, their economic expectations only deviate when they are primed with information about the economic benefits of the pandemic.

Even though the literature on priming in general and media priming in particular have been well-established, it is particularly interesting to study media priming during the Covid-19 pandemic. It is undoubtedly a "1 in 100 years" event in which media exposure is more intense than usual. Studying media priming within the specific situation of a pandemic brings about a number of valuable implications for policy making not only for the Covid-19 crisis but also for future extreme events. The fact that people hold a relatively high optimistic view on environmental quality amidst the Covid-19 pandemic is alarming. Governments and media should pay attention to avoid raising further optimism on how the disaster will affect the environment. Citizens may be otherwise less likely to support policies that prioritize environmental issues and to engage in mitigation behavior once the peak of the crisis will end. This scenario would constitute a serious threat to the achievement of the ambitious goals set by the COP21 and recently renovated by the European Green Deal. At the same time, the news could stress the positive possible economic benefits of the pandemic. Even such a short-lived priming like the one implemented here triggers optimism and probably releases some of the stress caused by the negative economic news. In turn, this can speed up countries' economic recovery.

Like all studies, our study has some limitations. We did not measure some key concepts that might bring more insights into the process of the media priming such as the use of media and how it has been changed during a critical period of time like the Covid-19 pandemic; prior expectations of individuals about the economic and environmental impact of the pandemic. An important venues to increase the generalizability of our results is to conduct the experiment on different target populations and to employ other priming techniques such as image or scenario priming.

*Psychology*, *34*, 156–168. https://doi.org/10.1016/j.joep.2012.09.007

George, J. M. (1991). State or trait: Effects of positive mood on prosocial behaviors at work. *Journal of applied Psychology*, *76*(2), 299.

Goff, S. H., Waring, T. M., and Noblet, C. L. (2017). Does Pricing Nature Reduce Monetary Support for Conservation?: Evidence From Donation Behavior in an Online Experiment. *Ecological Economics*, *141*, 119–126. https://doi.org/10.1016/j.ecolecon.2017.05.027

Hanley, N., Boyce, C., Czajkowski, M., Tucker, S., Noussair, C., and Townsend, M. (2017). Sad or Happy? The Effects of Emotions on Stated Preferences for Environmental Goods. *Environmental and Resource Economics*, *68*(4), 821–846. https://doi.org/10.1007/s10640-016-0048-9

Helm, D. (2020). The environmental impacts of the Coronavirus. *Environmental and Resource Economics*, (26), 1–20. https://doi.org/10.1007/s10640-020-00426-z

Henrich, J., Heine, S. J., and Norenzayan, A. (2010). The weirdest people in the world? *Behavioral and Brain Sciences*, *33*(2–3), 61–83. https://doi.org/10.1017/S0140525X0999152X

Henrique, K. P., and Tschakert, P. (2019). Taming São Paulo's floods: Dominant discourses, exclusionary practices, and the complicity of the media. *Global Environmental Change*, *58*, 101940. https://doi.org/10.1016/j.gloenvcha.2019.101940

Hollanders, D., and Vliegenthart, R. (2011). The influence of negative newspaper coverage on consumer confidence: The Dutch case. *Journal of Economic Psychology*, *32*(3), 367–373. https://doi.org/10.1016/j.joep.2011.01.003

Horton, J. J., Rand, D. G., and Zeckhauser, R. J. (2011). The online laboratory: Conducting experiments in a real labor market. *Experimental Economics*, *14*(3), 399–425. https://doi.org/10.1007/s10683-011-9273-9

Janiszewski, C., and Wyer, R. S. (2014). Content and process priming: A review. *Journal of Consumer Psychology*. https://doi.org/10.1016/j.jcps.2013.05.006

Kalogeropoulos, A., Albæk, E., De Vreese, C. H., and Van Dalen, A. (2017). News priming and the changing economy: How economic news influences government evaluations. *International Journal of Public Opinion Research*. https://doi.org/10.1093/ijpor/edv048

Kamenica, E. (2012). Behavioral Economics and Psychology of Incentives. *Annual Review of Economics*, *4*(1), 427–452. https://doi.org/10.1146/annurev-economics-080511-110909

Keizer, K., Lindenberg, S., and Steg, L. (2008). The Spreading of Disorder. *Science*, *322*(5908), 1681–1685. https://doi.org/10.1126/science.1161405

Larney, A., Rotella, A., and Barclay, P. (2019). Stake size effects in ultimatum game and dictator game offers: A meta-analysis. *Organizational Behavior and Human Decision Processes*. https://doi.org/10.1016/j.obhdp.2019.01.002

Lei, H., Yuting, H., Jun, B., Ying, Z., Yang, L., and Hammitt, J. K. (2013). Effect of the Fukushima nuclear accident on the risk perception of residents near a nuclear power plant in China. *Proceedings of the*
27